\documentclass{article}

\usepackage{authblk}
\usepackage[english]{babel}
\usepackage[utf8x]{inputenc}
\usepackage{amsmath}
\usepackage{graphicx, subcaption}
\usepackage[colorinlistoftodos]{todonotes}
\usepackage{float}
\usepackage{hyperref}
\usepackage{tikz}
\usepackage{cite}
\begin{document}
\title{Perturbations and quantum relaxation}
\author{Adithya Kandhadai, Antony Valentini}
\affil{\textit{Department of Physics and Astronomy, Clemson University, Kinard Laboratory, Clemson, SC 29634-0978, USA}}
\date{}

\maketitle

\begin{abstract}
We investigate whether small perturbations can cause relaxation to quantum equilibrium over very long timescales. We consider in particular a two-dimensional harmonic oscillator, which can serve as a model of a field mode on expanding space. We assume an initial wave function with small perturbations to the ground state. We present evidence that the trajectories are highly confined so as to preclude relaxation to equilibrium even over very long timescales. Cosmological implications are briefly discussed.
\end{abstract}

\section{Introduction}

The de Broglie-Bohm pilot-wave formulation of quantum theory
\cite{deB28, BV09, B52a, B52b, Holl93} provides a generalisation of the quantum
formalism, in which probabilities may differ from those predicted by the usual
Born rule \cite{AV91a, AV91b, AV92, AV96, AV01, PV06}. The Born rule applies only
to a statistical state of quantum equilibrium, which may be understood as
arising from a process of dynamical relaxation or `quantum relaxation'
(analogous to thermal relaxation)
\cite{AV91a, AV92, AV01, VW05, EC06, TRV12, SC12, ACV14}. Quantum nonequilibrium may
have existed in the early universe \cite{AV91a, AV91b, AV92, AV96, AV01, AV09}, in
which case violations of the Born rule could leave discernible traces in the
cosmic microwave background (CMB) \cite{AV07, AV10, CV13, CV15, AV15, CV16} and
perhaps even survive until today for certain relic cosmological particles
\cite{AV01, AV07, AV08, UV15, UV16}. Our current understanding of these cosmological
scenarios depends, however, on our understanding of quantum relaxation --
which remains incomplete in some important respects. In particular, the effect
of small perturbations has not been considered. As we shall see, from a
cosmological point of view it is important to establish what the effect of
small perturbations might be, in particular over long timescales.

In pilot-wave theory, a system has a definite configuration $q(t)$ which
evolves in time according to a law of motion for its velocity, where $\dot
{q}\equiv dq/dt$ is determined by the wave function $\psi(q,t)$. Here $\psi$
satisfies the usual Schr\"{o}dinger equation $i\partial\psi/\partial t=\hat
{H}\psi$ (taking $\hbar=1$). For standard Hamiltonians, $\dot{q}$ is
proportional to the gradient $\partial_{q}S$ of the phase $S$ of $\psi$. More
generally, $\dot{q}=j/|\psi|^{2}$ where $j=j\left[  \psi\right]  =j(q,t)$ is
the Schr\"{o}dinger current \cite{SV08}. The `pilot wave' $\psi$ guides the
motion of an individual system and in principle has no connection with
probability. For an ensemble of systems with the same wave function, we may
consider an arbitrary initial distribution $\rho(q,0)$ (at $t=0$) of
configurations $q(0)$. By construction, the time evolution $\rho(q,t)$ will
obey the continuity equation%
\begin{equation}
\frac{\partial\rho}{\partial t}+\partial_{q}\cdot\left(  \rho\dot{q}\right)
=0\ .
\end{equation}
Because $\left\vert \psi\right\vert ^{2}$ obeys the same equation, an initial
distribution $\rho(q,0)=\left\vert \psi(q,0)\right\vert ^{2}$ will evolve into
$\rho(q,t)=\left\vert \psi(q,t)\right\vert ^{2}$. In this equilibrium state,
probabilities match the Born rule and pilot-wave theory reproduces the usual
predictions of quantum theory \cite{B52a,B52b}. But we may just as well
consider nonequilibrium distributions $\rho(q,0)\neq\left\vert \psi
(q,0)\right\vert ^{2}$, opening up the possibility of a new and wider physics
with violations of the Born rule and new phenomena outside the domain of
conventional quantum physics \cite{AV91a, AV91b, AV92, AV96, AV01, AV02, AV07, AV08, AV08a, AV09, AV10, AVPwtMw, PV06}.\footnote{Note that the properties of a general nonequilibrium ensemble cannot be described by $\psi$ alone.}

Quantum relaxation to the equilibrium state $\rho=\left\vert \psi\right\vert
^{2}$ may be quantified by a coarse-grained $H$-function%
\begin{equation}
\bar{H}=\int dq\ \bar{\rho}\ln(\bar{\rho}/\overline{\left\vert \psi\right\vert
^{2}})\ ,
\end{equation}
where $\bar{\rho}$, $\overline{\left\vert \psi\right\vert ^{2}}$ are obtained
by coarse-graining $\rho$, $\left\vert \psi\right\vert ^{2}$ respectively.
This obeys a coarse-graining $H$-theorem $\bar{H}(t)\leq\bar{H}(0)$ (if the
initial state has no fine-grained micro-structure) \cite{AV91a,AV92,AV01}. The
minimum $\bar{H}=0$ corresponds to equilibrium $\bar{\rho}=\overline
{\left\vert \psi\right\vert ^{2}}$. While this provides some understanding of
how equilibrium is approached, the extent of relaxation depends on the system
and on the initial conditions. For two-dimensional systems with wave functions
that are evenly-weighted superpositions of energy eigenstates, extensive
numerical studies have shown that initial nonequilibrium distributions $\rho$
(with no fine-grained micro-structure) rapidly approach $\left\vert
\psi\right\vert ^{2}$ on a coarse-grained level
\cite{AV92,AV01,VW05,TRV12,SC12}, with $\bar{H}(t)$ decaying approximately
exponentially with time \cite{VW05,TRV12}. In these examples, the wave
function is periodic in time and the simulations were carried out up to one
period $T$. More recently, such simulations were extended to longer timescales
(up to $50T$) \cite{ACV14}. It was found that, for some initial wave functions
(with certain choices of initial phases), the decay of $\bar{H}(t)$ saturates
to a small but non-zero residue -- signalling an incomplete relaxation. This
was shown to occur when a significant fraction of the trajectories remain
confined to sub-regions and do not explore the full support of $\left\vert
\psi\right\vert ^{2}$. The numerical evidence indicated that such confinement
(and the associated incomplete relaxation) is less likely to occur for larger
numbers of superposed energy states \cite{ACV14}. These conclusions are
consistent with earlier examples studied by Colin \cite{SC12} and by
Contopoulos \textit{et al}. \cite{CDE12}, in which limited relaxation -- and
an associated confinement of trajectories -- was found for some initial wave
functions with only three or four energy states.

Previous studies of quantum relaxation have mostly focussed on a
coarse-graining approach for isolated systems \cite{AV91a,AV92,AV01}, modelled
on the analogous classical discussion \cite{Tol, Dav}.\footnote{An exception
is an early paper by Bohm \cite{B53}, which considered an ensemble of
two-level molecules subject to random external collisions and argued that the
molecules would relax to equilibrium.} In this paper we consider instead the
effect of small perturbations, in particular over very long timescales (of
order $10^{3}T$). This is of interest in its own right, as well as for
cosmological reasons.

Consider a system with an unperturbed wave function $\psi$, which generates an
unperturbed velocity field $\dot{q}$ and unperturbed trajectories $q(t)$. The
system might be subjected to small external perturbations, which in a first
approximation we may model as perturbations to the classical potential of the
system. The system will then have a perturbed wave function $\psi^{\prime}$
which is close to $\psi$, and a perturbed velocity field $\dot{q}^{\prime}$
which we expect to be close to $\dot{q}$. Will the perturbed trajectories
$q^{\prime}(t)$ remain close to $q(t)$? One might expect that even a small
difference in the velocity field, acting over sufficiently long periods of
time, would yield perturbed trajectories $q^{\prime}(t)$ which deviate greatly
from $q(t)$. For example, one might consider a two-dimensional harmonic
oscillator with configuration $q=(q_{1},q_{2})$ whose unperturbed wave
function is simply the ground state, $\psi(q_{1},q_{2},t)=\phi_{0}(q_{1}%
)\phi_{0}(q_{2})e^{-iE_{0}t}$, where $\phi_{0}(q_{1})\phi_{0}(q_{2})$ is a
real Gaussian and $E_{0}$ is the ground-state energy. Because the phase
$S=\operatorname{Im}\ln\psi$ is independent of position, the unperturbed
velocity field $\dot{q}$ vanishes everywhere and all unperturbed trajectories
are static. There can be no relaxation, nor indeed any evolution at all of the
unperturbed density $\rho$. Any initial nonequilibrium distribution
$\rho(q_{1},q_{2},0)\neq\left\vert \phi_{0}(q_{1})\phi_{0}(q_{2})\right\vert
^{2}$ will remain the same. Now let us consider a perturbed wave function
$\psi^{\prime}$ that differs from $\psi$ by the addition of excited states
$\phi_{m}(q_{1})\phi_{n}(q_{2})$ with small amplitudes $\epsilon_{mn}$. For
small $\epsilon_{mn}$ the perturbed velocity field $\dot{q}^{\prime}$ will be
small but generally non-zero. The question is: over arbitrarily long times,
will the perturbed trajectories $q^{\prime}(t)$ remain confined to small
sub-regions of the support of $\left\vert \phi_{0}(q_{1})\phi_{0}%
(q_{2})\right\vert ^{2}$ or will they wander over larger regions and possibly
over the bulk of the support of $\left\vert \phi_{0}(q_{1})\phi_{0}%
(q_{2})\right\vert ^{2}$? In the former case, there could be no relaxation
even over arbitrarily long times. In the latter case, relaxation could occur.
Indeed, in the latter case it might seem plausible that, no matter how small
$\epsilon_{mn}$ may be, over sufficiently long timescales the perturbed
distribution $\rho^{\prime}(q_{1},q_{2},t)$ could approach $\left\vert
\phi_{0}(q_{1})\phi_{0}(q_{2})\right\vert ^{2}$ to arbitrary accuracy (where
$\left\vert \phi_{0}(q_{1})\phi_{0}(q_{2})\right\vert ^{2}$ coincides with
equilibrium as $\epsilon_{mn}\rightarrow0$). The question, then, is whether
small perturbations are generally ineffective for relaxation or whether they might
conceivably drive systems to equilibrium over sufficiently long times.

Cosmologically, the effect of perturbations over long timescales could be
important for several reasons. According to inflationary cosmology, the
temperature anisotropies in the CMB were seeded by primordial quantum
fluctuations of a scalar field whose quantum state was approximately a vacuum
(the Bunch-Davies vacuum) \cite{LL00, Muk05, PU09}. It has been shown that de
Broglie-Bohm trajectories for field amplitudes in the Bunch-Davies vacuum are
too trivial to allow relaxation \cite{AV07, AV10}. On this basis it was
concluded that, if quantum nonequilibrium existed at the beginning of
inflation, then it would persist throughout the inflationary phase and
potentially leave an observable imprint in the CMB today. However, strictly
speaking this conclusion depends on the implicit assumption that (unavoidable)
small corrections to the Bunch-Davies vacuum can be neglected in the sense
that they will not generate relaxation during the inflationary era. Similarly,
a cosmological scenario has been developed according to which quantum
relaxation occurred during a pre-inflationary (radiation-dominated) phase
\cite{AV07, AV10, CV13, CV15, AV08, CV16}. It was shown that during such a phase
relaxation proceeds efficiently at short wavelengths but is suppressed at long
wavelengths, resulting in a distinctive signature of quantum nonequilibrium at
the beginning of inflation -- which is then imprinted at later times in the
CMB.\footnote{Specifically, the signature amounts to a primordial power
deficit at long wavelengths with a specific (inverse-tangent) dependence on
wavenumber $k$ \cite{CV15,CV16}. A large-scale power deficit has in fact been
reported in the \textit{Planck} data
\cite{PlanckXV-2013,Planck15-XI-PowerSpec}, though the extent to which it
matches our prediction is still being evaluated \cite{VVP16}.} The resulting
predictions for the CMB depend, however, on the assertion that there will be
no significant relaxation during inflation itself, an assertion which again
depends on the implicit assumption that small corrections to the Bunch-Davies
vacuum may be ignored. Finally, a scenario has also been developed according
to which, for certain particles created in the early universe, any
nonequilibrium carried by them could conceivably survive (or partially
survive) to the present \cite{UV15}. But such a scenario would fail if small
perturbations caused the particles to relax over very long timescales. Indeed,
if small perturbations do cause relaxation over long timescales it would be
exceedingly difficult to have any hope at all of discovering relic
nonequilibrium today.

To discuss these cosmological matters quantitatively, it suffices to consider
a free (minimally-coupled) massless scalar field $\phi$ on expanding flat
space with scale factor $a(t)$. Here $t$ is standard cosmological time and
physical wavelengths are proportional to $a(t)$. In Fourier space we have
field components $\phi_{\mathbf{k}}(t)$ which may be written in terms of their
real and imaginary parts, $\phi_{\mathbf{k}}=\frac{\sqrt{V}}{(2\pi)^{3/2}%
}\left(  q_{\mathbf{k}1}+iq_{\mathbf{k}2}\right)  $ (where $V$ is a
normalisation volume). The field Hamiltonian then becomes a sum $H=\sum
_{\mathbf{k}r}H_{\mathbf{k}r}$, where $H_{\mathbf{k}r}$ ($r=1,2$) is
mathematically the Hamiltonian of a harmonic oscillator with mass $m=a^{3}$
and angular frequency $\omega=k/a$ \cite{AV07,AV08,AV10}. If we consider an
unentangled mode $\mathbf{k}$, we have an independent dynamics with a wave
function $\psi=\psi(q_{1},q_{2},t)$ (dropping the index $\mathbf{k}$) that
satisfies the Schr\"{o}dinger equation%
\begin{equation}
i\frac{\partial\psi}{\partial t}=\sum_{r=1,\ 2}\left(  -\frac{1}{2m}%
\partial_{r}^{2}+\frac{1}{2}m\omega^{2}q_{r}^{2}\right)  \psi
\end{equation}
(with $\partial_{r}\equiv\partial/\partial q_{r}$). The pilot-wave equation of
motion for the actual configuration $(q_{1},q_{2})$ then reads%
\begin{equation}
\dot{q}_{r}=\frac{1}{m}\operatorname{Im}\frac{\partial_{r}\psi}{\psi
}\label{deB}%
\end{equation}
and an arbitrary marginal distribution $\rho=\rho(q_{1},q_{2},t)$ will then
evolve according to the continuity equation%
\begin{equation}
\frac{\partial\rho}{\partial t}+\sum_{r=1,\ 2}\partial_{r}\left(  \rho\frac
{1}{m}\operatorname{Im}\frac{\partial_{r}\psi}{\psi}\right)  =0\ .
\end{equation}
These equations are just those of pilot-wave dynamics for a two-dimensional
harmonic oscillator with (time-dependent) mass $m=a^{3}$ and angular frequency
$\omega=k/a$. It may be shown that the resulting time evolution is
mathematically equivalent to that of an ordinary harmonic oscillator (with
constant mass and angular frequency) but with the time parameter replaced by a
`retarded time' that depends on $k$ \cite{CV13}. It is found, in particular,
that relaxation is suppressed\ at long (super-Hubble) wavelengths while
proceeding efficiently at short (sub-Hubble) wavelengths \cite{AV08,CV13,CV15}.

Thus cosmological relaxation for a single field mode may be discussed in terms
of relaxation for a standard oscillator. By studying the effect of small
perturbations on relaxation for a simple two-dimensional harmonic oscillator,
then, we may draw conclusions that have application to cosmology.

Before proceeding, we may comment on an alternative approach to understanding the Born rule in pilot-wave theory. Beginning with ref. \cite{DGZ92}, it has been argued that the Born rule is obtained for subsystems in a `typical' universe (see also ref. \cite{DT09}). However, as noted in refs. \cite{AV96, AV01}, this conclusion depends on defining `typicality' (or probability) with respect to the Born-rule measure $|\psi|^2$ where $\psi$ is the wave function for the whole universe. Other choices, such as $|\psi|^4$, yield different typical distributions for subsystems. Thus quantum equilibrium for subsystems is indeed `typical' (has unit measure) with respect to the global Born-rule measure $|\psi|^2$; but equally it is `untypical' (has vanishing measure) with respect to a global non-Born-rule measure. In effect, in this alternative approach it is assumed that initial conditions are chosen randomly with respect to $|\psi|^2$. We see no reason to assume this. Pilot-wave dynamics is a deterministic theory whose initial conditions are in principle arbitrary. The question of what the initial conditions for our universe actually were, and in particular of whether subsystems were initially in quantum equilibrium or not, is ultimately an empirical question that can be settled only by cosmological or astrophysical observations \cite{AV01, AV10, UV15, CV15}.

Generally speaking, as argued in ref. \cite{AV01}, there is an inevitable empirical component to reasoning in statistical mechanics. In the case of pilot-wave theory, from the observation of approximate equilibrium today -- to present experimental accuracy, and for the systems that have been probed so far -- we may deduce that past conditions must have been such as to evolve to approximate equilibrium today to the said accuracy and for the said systems. We may then focus our investigations on initial conditions that satisfy these basic criteria. Thus we may consider initial conditions that evolve to nonequilibrium today, provided the nonequilibrium is either too small to have been detected so far or occurs in exotic systems that have not yet been probed. Previous studies and simulations have revealed a broad class of possibilities \cite{VW05, EC06, TRV12, SC12, ACV14, AV07, AV10, CV13, CV15, AV15, CV16, AV08, UV15, UV16}. A question studied in this paper is whether a scenario is possible in which the present approximate equilibrium was brought about by past small perturbations only (acting over very long timescales). Our results suggest that this is not possible. At the same time, our results make it more plausible that relic nonequilibrium from early times could survive over cosmological timescales and possibly still exist today.

It might be thought that our results perhaps follow trivially from the classical KAM theorem, which concerns the effect of small perturbations on Hamiltonian systems. But in fact the KAM theorem is of no use here. Pilot-wave dynamics is very different from classical mechanics, in particular because it is a first-order dynamics in configuration space not in phase space. Pilot-wave theory can be rewritten in a Hamiltonian form but at the price of introducing (a) a constraint $p = \partial_q S$ on the momenta $p$ at some initial time $t = 0$ and (b) introducing an additional potential $Q$ (the `quantum potential') which depends on $\psi$ and which is generally time-dependent. It has been shown in ref. \cite{CV2014} that the requirement (a) renders the theory unstable, so the Hamiltonian reformulation is not physically acceptable. Furthermore, even if one does base the dynamics on Hamilton's equations by artificially restricting the theory to the exact surface $p = \partial_q S$ in phase space, the standard KAM theorem will not apply because the extra potential $Q$ is time-dependent. Finally we note that, in any case, the KAM theorem concerns the effect of small perturbations on trajectories in phase space whereas our work concerns the effect of small perturbations on trajectories in configuration space.

To understand the effect of perturbations in this very different form of dynamics therefore requires us to consider the matter afresh. Our approach employs numerical experiments to gather evidence for our conclusion. A theoretical understanding of our results is left for future work.

In Section 2 we present our model, which is obtained simply by setting $m=1$
in the equation of motion (\ref{deB}) (for $r=1,2$). This defines our dynamics
of trajectories for a two-dimensional harmonic oscillator, with constant mass
and constant angular frequency and with a given wave function. We shall take
$\psi$ to be the ground state with small perturbations of amplitude
$\epsilon_{mn}$ coming from the lowest excited states $\phi_{m}(q_{1})\phi
_{n}(q_{2})$. In Section 3 we discuss our method, where two different techniques are applied to infer the extent of relaxation in the long-time limit,  using samples of trajectories evolved over long times. In
Section 4 we then study numerically the behaviour of a sample of trajectories
over very long timescales, in particular we consider how their behaviour
changes as the perturbations become smaller. As we shall see, for sufficiently
small perturbations the trajectories become highly confined, and neighboring trajectories are confined to almost the same regions, even over very
long timescales -- from which we conclude (tentatively, given our numerical
evidence) that small perturbations do not cause relaxation. Cosmological
implications are briefly discussed in Section 5, where we draw our conclusions.

\section{Oscillator model}

The system under consideration is the standard two-dimensional harmonic oscillator. We employ units such that $\hbar = m = \omega = 1$. The wave function at $t = 0$ is taken to be the ground state of the oscillator perturbed by a superposition of excited states:

\begin{equation}
\psi(q_1, q_2, 0) = N\left[e^{i\theta_{00}}\phi_0(q_1)\phi_0(q_2) + \sum_{(m,n) \neq (0,0)} \epsilon_{mn}e^{i\theta_{mn}}\phi_m(q_1)\phi_n(q_2) \right] .
\end{equation}
Here $N$ is a suitable normalization factor.

The $\theta_{mn}$ are randomly chosen initial phases taking values between $0$ and $2\pi$. The function $\phi_m(q_r)$ is the eigenfunction corresponding to the $m^{th}$ energy state of the harmonic oscillator, given by

\begin{equation}
\phi_m(q_r) = \frac{1}{\pi^{\frac{1}{4}}} \frac{1}{\sqrt{2^m m!}} H_m(q_r) e^{-\frac{q_r^2}{2}} ,
\end{equation}
where $H_m$ is the Hermite polynomial of order $m$. For an energy eigenstate $\phi_m(q_1)\phi_n(q_2)$ of the two-dimensional harmonic oscillator, the corresponding energy eigenvalue is $E_{mn} = (m + n + 1)$ (in our units).

The parameters $\epsilon_{mn}$ take values between $0$ and $1$ and quantify the difference between the initial wave function and the ground state. For $\epsilon_{mn} = 1$, the initial wave function is an equally-weighted superposition of the first four energy states, as studied in \cite{ACV14}. For small values of $\epsilon_{mn}$, the initial wave function can be thought of as the ground state with small perturbations. 

Introducing the quantities $ \alpha_{mn}(t) = \theta_{mn} - E_{mn}t $, the wave function at any time $t$ is given by

\begin{equation}
\psi(q_1, q_2, t) = N\left[e^{i\alpha_{00}(t)}\phi_0(q_1)\phi_0(q_2) + \sum_{(m,n) \neq (0,0)} \epsilon_{mn}e^{i\alpha_{mn}(t)}\phi_m(q_1)\phi_n(q_2) \right] .
\end{equation}
Note that this wave function is periodic with period $T = 2\pi$.

The velocity field for this wave function is given by (4) (with $m = 1$), which is the equation of motion that determines our trajectories.

In most of our simulations, we shall for simplicity assume that $\epsilon_{mn} = \epsilon$, that is, we assume a `homogeneous' perturbation. Our wave function at time $t$ then reads

\begin{equation}
\psi(q_1, q_2, t) = N\left[e^{i\alpha_{00}}\phi_0(q_1)\phi_0(q_2) + \epsilon \sum_{(m,n) \neq (0,0)} e^{i\alpha_{mn}}\phi_m(q_1)\phi_n(q_2) \right] .
\end{equation}
We shall however verify with examples that similar results are obtained for unequal values of $\epsilon_{mn}$, so that this simplification is unimportant.

If the wave function is simply one of the energy eigenstates, as opposed to a superposition, then the trajectories are stationary (since the eigenstates of the harmonic oscillator are purely real, apart from an overall phase factor) and no relaxation occurs. However, a superposition of energy eigenstates will usually generate non-trivial trajectories and some degree of relaxation. The question is: if the perturbations are small, will significant relaxation still occur -- at least over sufficiently long timescales?

\section{Method}

As we have noted, our goal is to study the extent of relaxation for harmonic oscillator wave functions of the form (8) consisting of the ground state with small perturbations from excited states. We will be considering long timescales of order 1000 periods. It would be computationally intractable to simulate the evolution of a complete non-equilibrium distribution $\rho$ over such long times. For example, previous simulations of relaxation were carried out for up to 50 periods only \cite{ACV14}, which was already computationally demanding. It is therefore not possible to calculate the time evolution $\bar{H}(t)$ of the coarse-grained $H$-function, as would be required to quantify relaxation precisely. Instead, for each initial wave function, we shall begin by examining a sample of ten individual trajectories that start at the following points: $(q_1, q_2)$ = (1.5, 1.5), (1.5, -1.5), (-1.5, 1.5), (-1.5, -1.5), (0.5, 0.0), (0.0, -0.5), (-0.5, 0.0), (0.0, 0.5), (0.25, 0.25), and (0.25, -0.25). These points will be referred to as points 1 through 10, respectively. For a given initial wave function, the number of trajectories that travel over the main support of $|\psi|^2$ (as opposed to remaining confined to a small sub-region) and the degree to which they cover it may be used to draw preliminary inferences about the extent of relaxation. 

The justification for this method comes from ref. \cite{ACV14}, in which it was shown that limits on relaxation -- quantified by a non-zero `residue' of $\bar{H}(t)$ at large times -- occur when a significant fraction of the trajectories show a substantial degree of confinement (to a sub-region of the support of $|\psi|^2$). As is to be expected, confinement of trajectories is associated with a lack of complete relaxation (even over long timescales), while if the trajectories tend to wander over the bulk of the support of $|\psi|^2$ then a more complete relaxation can take place. For example an initial wave function with four modes, whose trajectories are analyzed in detail for the $\epsilon = 1$ case in the next section, was considered in ref. \cite{ACV14} and the chosen trajectories showed strong confinement. The same wave function yielded a significant non-zero residue in the long-time plot of $\bar{H}(t)$. In contrast, an initial wave function with 25 modes showed efficient relaxation with a vanishing residue (to within errors) for $\bar{H}(t)$, while its associated trajectories traveled over the bulk of the support of $|\psi|^2$.

There is of course no sharp dividing line between trajectories that are confined and those that are not, and we do not use a precise quantitative criterion. As in ref. \cite{ACV14} it will suffice to use our judgment in deciding how well the trajectories cover the space. The effectiveness of this method -- at least for present purposes -- will become clear in what follows. The advantage, in particular, is that we may deduce whether or not $\bar{H}(t)$ decays without actually having to calculate $\bar{H}(t)$ -- which would be completely intractable over the very long timescales considered here.

We shall also use another method, again inspired by ref. \cite{ACV14}, to infer whether or not relaxation to equilibrium occurs over very long timescales. We shall study trajectories starting from neighbouring initial points to see whether or not they are confined to essentially the same sub-regions. If they are so confined, this will constitute further evidence against relaxation. Examples of a similar kind of correspondence were reported in Section 5 of ref. \cite{ACV14}.

To plot a trajectory, we consider a particle starting at the required point and calculate its position every hundredth of a period. We employ a Dormand-Prince adaptive time step algorithm to solve the equation of motion (4) with the wave function (9). The algorithm has an upper bound on the allowable error in each time step (denoted ABSTOL) which is used to choose the step size. Our trajectories are accurate up to an absolute error of $0.01$ in the final position. This is confirmed by checking that the final positions with two consecutive values of ABSTOL are not separated by a distance greater than $0.01$.

\section{Behaviour of trajectories over very long timescales}

We now present numerical results for the behaviour of trajectories over very long timescales. We first consider the previously-discussed case $\epsilon$ = 1 with four modes\cite{ACV14}, though now over much longer timescales. We are able to confirm that the behaviour found in \cite{ACV14} persists over timescales about two orders of magnitude larger than were previously considered. We then examine how the behaviour changes for smaller $\epsilon$. Our results indicate that relaxation is suppressed for small $\epsilon$. We present evidence that small perturbations are unlikely to yield significant relaxation even over arbitrarily long timescales.

\pagebreak

\subsection{Case $\epsilon = 1$}

We first consider the same four-mode wave function, with $m$, $n$ in equation (9) summing from $0$ to $1$, that was investigated in \cite{ACV14}.\footnote{The initial phases were $\theta_{00} = 0.5442$, $\theta_{01} = 2.3099$, $\theta_{10} = 5.6703$, $\theta_{11} = 4.5333$.} In that work the confinement of the trajectories was studied up to 25 periods. The results are displayed in figure 8 of \cite{ACV14} and show that certain trajectories remain confined over 25 periods. (It was also found that the coarse-grained $H$-function seemed to saturate, with a small residue indicating incomplete relaxation -- caused by the confinement of a significant fraction of the trajectories.) We have now found that, for this same case, the confinement persists all the way up to 3000 periods. Initial points 1, 2, 3, 4 and 6 generate trajectories that wander over the bulk of the support of the wave function, while initial points 5, 7, 8, 9 and 10 generate trajectories that are confined to small sub-regions.

If a trajectory is found numerically to be confined to a sub-region over a given large time interval, it is of course conceivable that over even larger times the trajectory could wander outside of that region and possibly cover the bulk of the support of $|\psi|^2$. However, this is unlikely in the cases we have studied, for reasons that will be explained below.

To illustrate our results we may consider two specific trajectories, the first unconfined and the second confined.

\pagebreak

An example of an unconfined trajectory is shown in Fig. 1. The trajectory is plotted up to 25, 100, 200, 500, 1000 and 3000 periods. Clearly, the trajectory wanders over the bulk of the support of $|\psi|^2$ . (For comparison, the distribution $|\psi|^2$ is shown in Fig. 2.)
\begin{figure}[H]
    \begin{subfigure}{0.5\textwidth}
        \includegraphics[width = 4.5cm, height = 4cm]{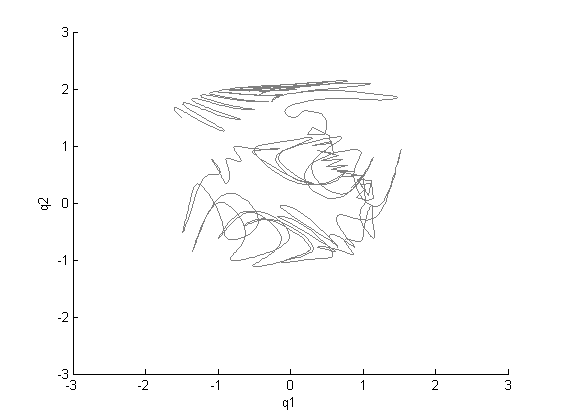}
        \subcaption{25T}
    \end{subfigure}
    \begin{subfigure}{0.5\textwidth}
        \includegraphics[width = 4.5cm, height = 4cm]{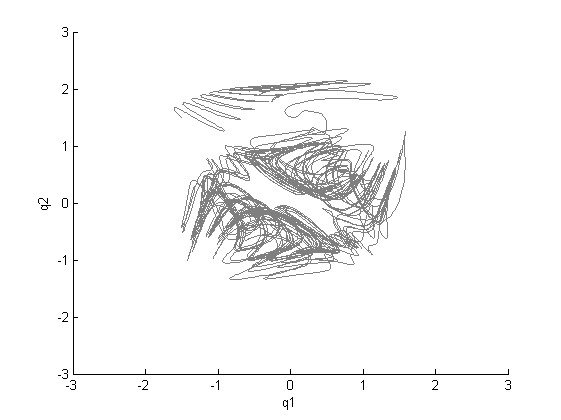}
        \subcaption{100T}
    \end{subfigure}
    \begin{subfigure}{0.5\textwidth}
        \includegraphics[width = 4.5cm, height = 4cm]{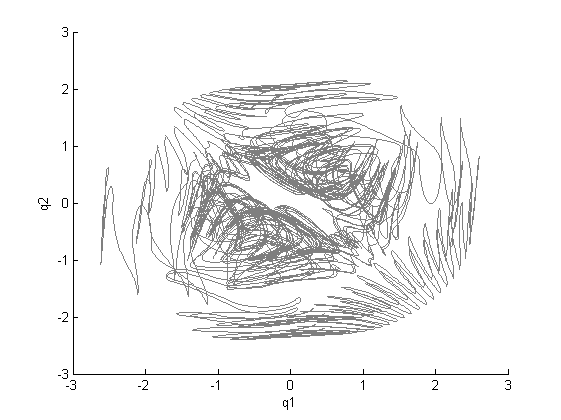}
        \subcaption{200T}
    \end{subfigure}
    \begin{subfigure}{0.5\textwidth}
        \includegraphics[width = 4.5cm, height = 4cm]{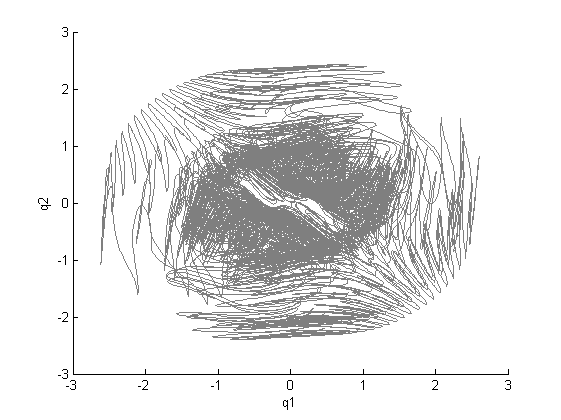}
        \subcaption{500T}
    \end{subfigure}
    \begin{subfigure}{0.5\textwidth}
        \includegraphics[width = 4.5cm, height = 4cm]{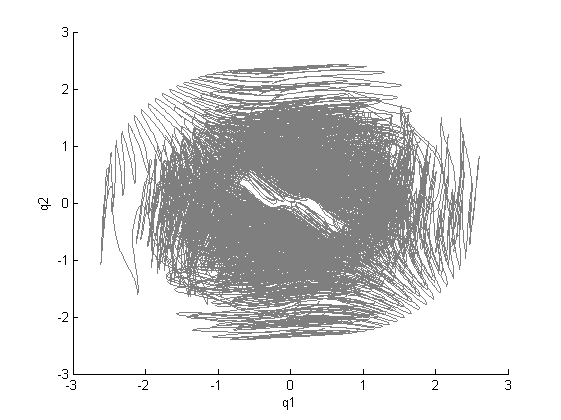}
        \subcaption{1000T}
    \end{subfigure}
    \begin{subfigure}{0.5\textwidth}
        \includegraphics[width = 4.5cm, height = 4cm]{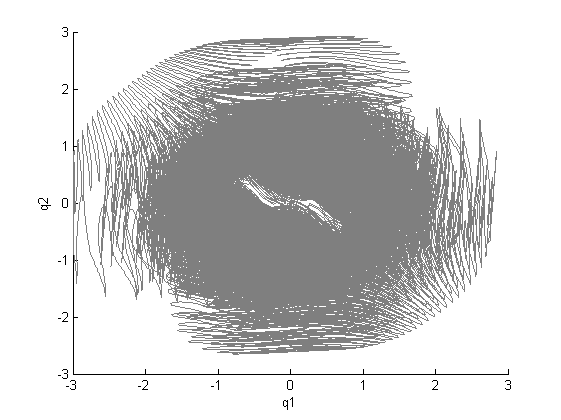}
        \subcaption{3000T}
    \end{subfigure}
     \caption{One of the unconfined trajectories for the wave function with initial phases specified in footnote 4, starting at the point (-1.5, 1.5). The six sub-figures are snapshots of the trajectory at 25, 100, 200, 500, 1000 and 3000 periods, as indicated below each.}
\end{figure}
\pagebreak

For comparison, the Born distribution is shown below in Fig. 2.

\begin{figure}[H]
\centering
\includegraphics[width=6cm, height=6cm]{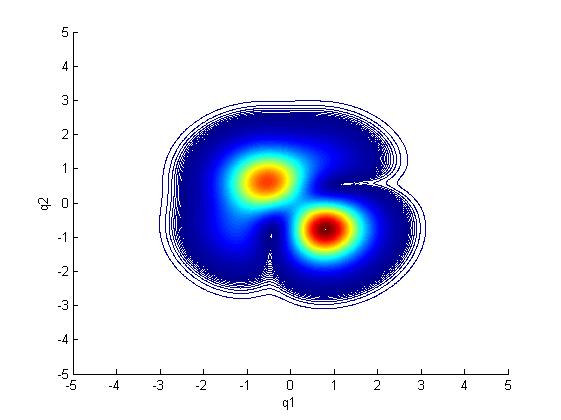}
\caption{The equilibrium distribution $|\psi|^2$ at the end of any integer number of periods.}
\end{figure}

The widths and heights of the trajectory in Fig. 1, calculated for the final
times 25, 100, 200, 500, 1000, 2000 and 3000 periods, are shown in Fig. 3. The
increasing range of the trajectory is obvious.
\begin{figure}[H]
    \begin{subfigure}{0.5\textwidth}
        \includegraphics[width = 4.5cm, height = 4cm]{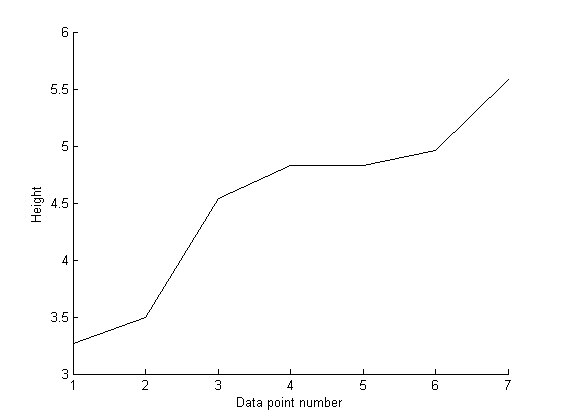}
    \end{subfigure}
    \begin{subfigure}{0.5\textwidth}
        \includegraphics[width = 4.5cm, height = 4cm]{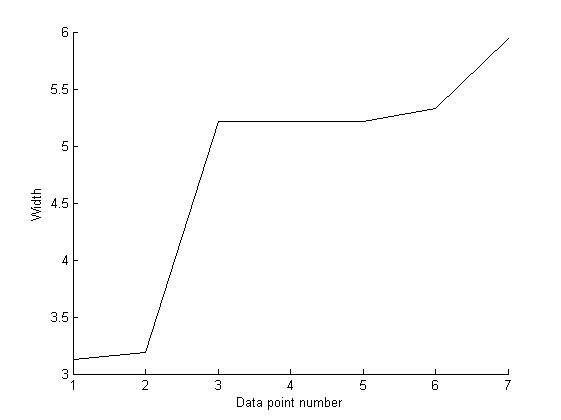}
    \end{subfigure}
    \caption{The height (a) and the width (b) of the unconfined
trajectory starting at (-1.5, 1.5). The data points are for the final times 25, 100, 200, 500, 1000, 2000 and 3000 periods.}
\end{figure}
\pagebreak

An example of a confined trajectory is shown in Fig. 4. The trajectory is again plotted up to 25, 100, 200, 500, 1000 and 3000 periods. The trajectory is clearly confined to a sub-region of the support of $|\psi|^2$ (compare Fig. 4 with Fig. 2).
\begin{figure}[H]
    \begin{subfigure}{0.5\textwidth}
        \includegraphics[width = 4.5cm, height = 4cm]{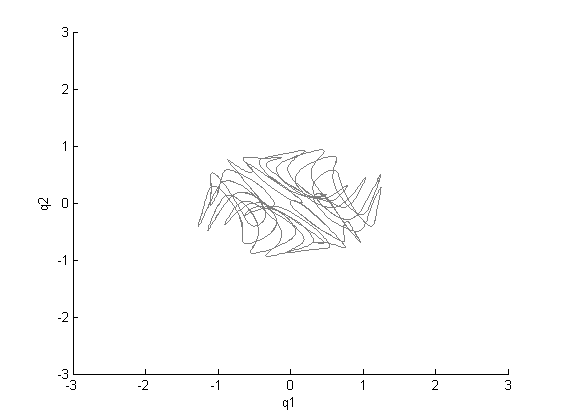}
        \subcaption{25T}
    \end{subfigure}
    \begin{subfigure}{0.5\textwidth}
        \includegraphics[width = 4.5cm, height = 4cm]{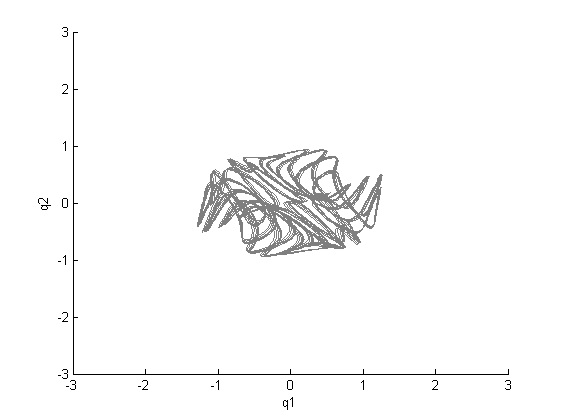}
        \subcaption{100T}
    \end{subfigure}
    \begin{subfigure}{0.5\textwidth}
        \includegraphics[width = 4.5cm, height = 4cm]{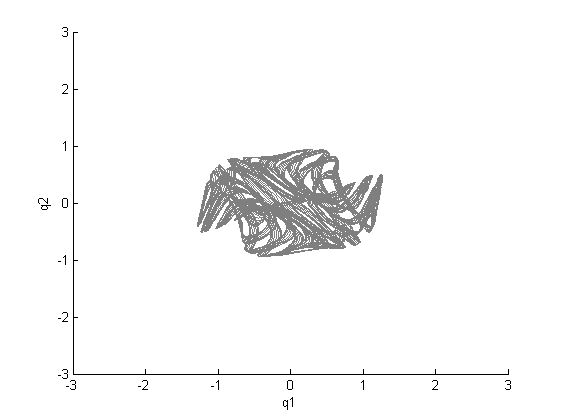}
        \subcaption{200T}
    \end{subfigure}
    \begin{subfigure}{0.5\textwidth}
        \includegraphics[width = 4.5cm, height = 4cm]{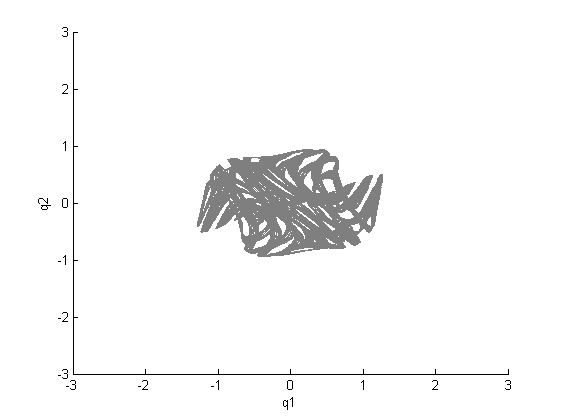}
        \subcaption{500T}
    \end{subfigure}
    \begin{subfigure}{0.5\textwidth}
        \includegraphics[width = 4.5cm, height = 4cm]{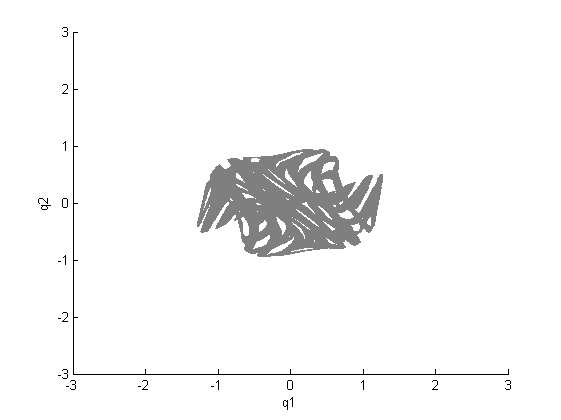}
        \subcaption{1000T}
    \end{subfigure}
    \begin{subfigure}{0.5\textwidth}
        \includegraphics[width = 4.5cm, height = 4cm]{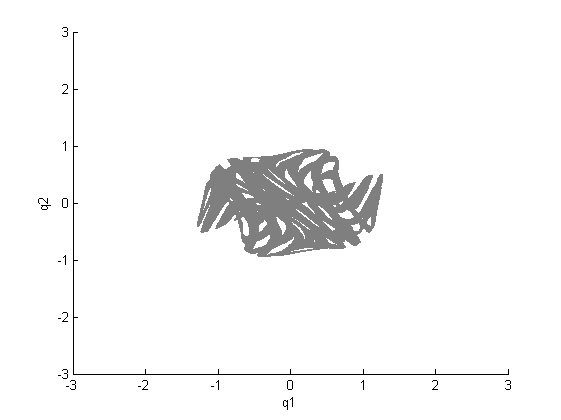}
        \subcaption{3000T}
    \end{subfigure}
     \caption{One of the confined trajectories for the same wave function as in Fig. 1, starting at (-0.5, 0.0). The six sub-figures are snapshots of the trajectory at 25, 100, 200, 500, 1000 and 3000 periods, as indicated below each.}
\end{figure}
\pagebreak

Evidence for strict confinement comes from plotting the width and height of the trajectory at the final times 25, 100, 200, 500, 1000, 2000 and 3000 periods. The results are shown in Fig. 5. There is a quick saturation after the early increase, indicating a strict and indefinite confinement to a sub-region (of the corresponding saturation width and height).

\begin{figure}[H]
    \begin{subfigure}{0.5\textwidth}
        \includegraphics[width = 4.5cm, height = 4cm]{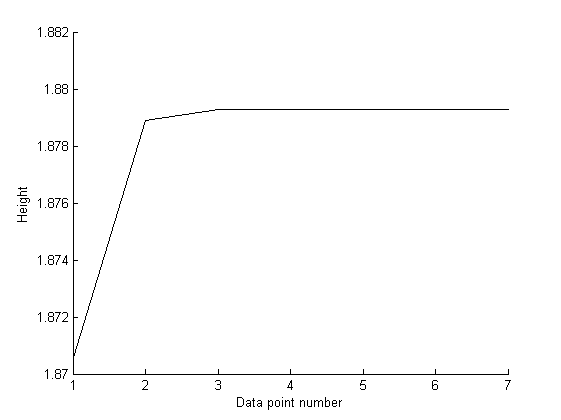}
    \end{subfigure}
    \begin{subfigure}{0.5\textwidth}
        \includegraphics[width = 4.5cm, height = 4cm]{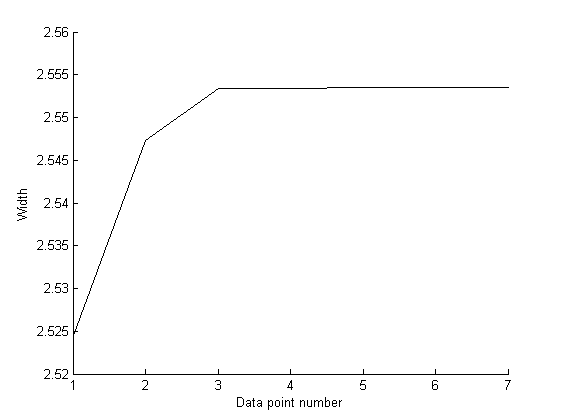}
    \end{subfigure}
    \caption{The height (a) and the width (b) of the confined
trajectory starting at (-0.5, 0). The data points are for the final times 25, 100, 200, 500, 1000, 2000 and 3000 periods.}
\end{figure}

To understand the overall behavior for $\epsilon = 1$, we have calculated trajectories starting at the same ten points 1-10 but for ten additional sets of randomly-generated initial phases $\theta_{mn}$ (hence a total of 100 additional trajectories). For about half of the initial wave functions, the trajectories are split more or less evenly between unconfined and confined (as was the case for the initial phases given in footnote 4), while for the remaining initial wave functions we find that 7-8 trajectories are unconfined and 2-3 are confined.

\subsection{Results for smaller $\epsilon$. Effect on confinement}

We wish to investigate how relaxation over very long timescales will be affected when $\epsilon$ is made smaller. We hope to find behavior that is largely independent of the choice of initial phases. To this end we generate trajectories with the same ten starting points 1-10 and for the same additional ten sets of values of $\theta_{mn}$ as in Section 4, but now with $\epsilon = 0.5$ and then again with $\epsilon = 0.25$, $0.1$ and $0.05$. \footnote{For the illustrative figures in this subsection, the initial phases were as follows: $\theta_{00} = 4.8157$, $\theta_{01} = 1.486$, $\theta_{10} = 2.6226$, $\theta_{11} = 3.8416$.}

For $\epsilon = 0.5$, it was almost always the case (for almost all choices of initial phases) that points 1-4 had largely unconfined trajectories while points 5-10 were confined to small sub-regions. Examples of both types of trajectory are shown in Fig. 6.

\begin{figure}[H]
    \begin{subfigure}{0.5\textwidth}
        \includegraphics[width = 6cm, height = 5cm]{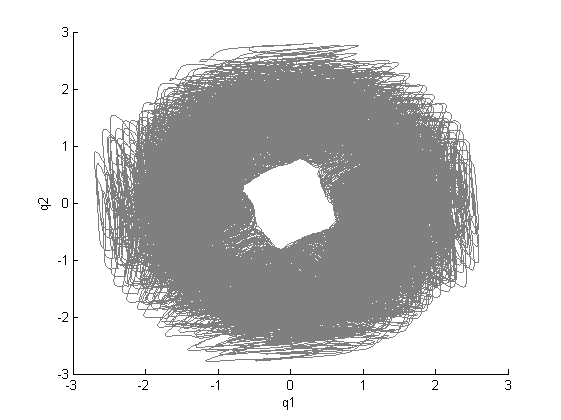}
        \subcaption{}
    \end{subfigure}
    \begin{subfigure}{0.5\textwidth}
        \includegraphics[width = 6cm, height = 5cm]{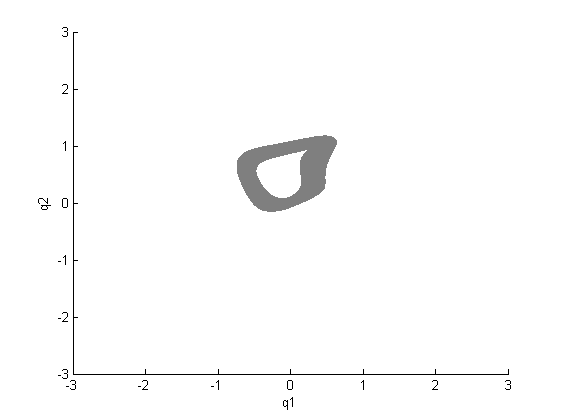}
        \subcaption{}
    \end{subfigure}
    \caption{Part (a) shows the trajectory starting from (-1.5, 1.5) plotted at the end of 3000 periods, for $\epsilon = 0.5$, with the set of initial phases listed in footnote 5. The trajectory is largely unconfined, though it does avoid a small central region. Part (b) shows the confined trajectory starting from (-0.5, 0), plotted at the end of 3000 periods, for the same wave function.}
\end{figure}
\pagebreak

When $\epsilon$ is decreased to $0.25$, a pattern begins to emerge. The initial points 1-4 orbit the origin while remaining confined to outer annular regions. The trajectories travel over large distances but leave a large empty space in the inner part of the support of $|\psi|^2$, and so may be considered confined. Points 5-10 are again confined to small sub-regions, as we saw for $\epsilon = 0.5$. Examples of both types of behaviour are shown in Fig. 7.

\begin{figure}[H]
    \begin{subfigure}{0.5\textwidth}
        \includegraphics[width = 6cm, height = 5cm]{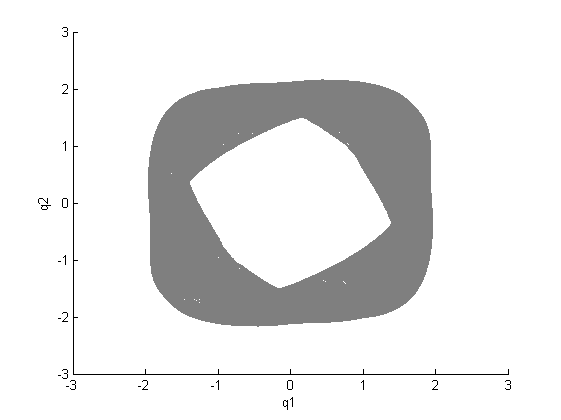}
        \subcaption{}
    \end{subfigure}
    \begin{subfigure}{0.5\textwidth}
        \includegraphics[width = 6cm, height = 5cm]{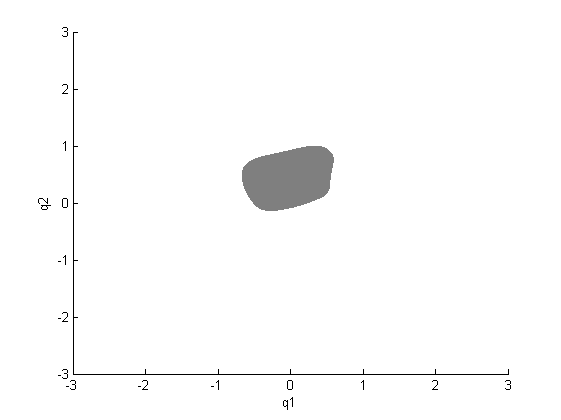}
        \subcaption{}
    \end{subfigure}
    \caption{Part (a) shows the trajectory starting from (1.5, 1.5) plotted at the end of 3000 periods, for $\epsilon = 0.25$, with the set of initial phases listed in footnote 5. The trajectory travels over a large distance, but avoids a significant fraction of the support of $|\psi|^2$. Part (b) shows the confined trajectory starting from (-0.5, 0) also plotted at the end of 3000 periods, for the same wave function.}
\end{figure}
\pagebreak
It is found that the motion of the trajectory in Fig. 7(a) may be decomposed into a quasi-oscillation confined to a small region together with an approximately uniform circular motion around the origin. This is shown in Fig. 8.

\begin{figure}[H]
    \begin{subfigure}{0.5\textwidth}
        \includegraphics[width = 4.5cm, height = 4cm]{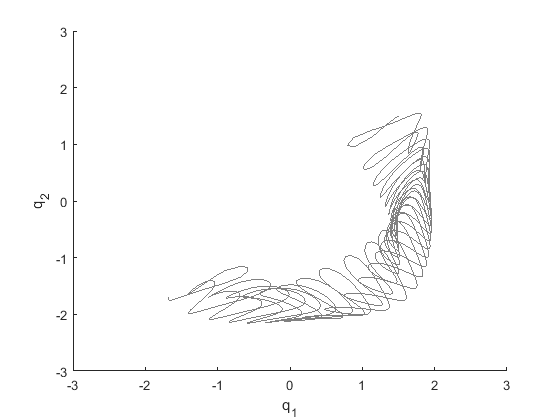}
        \subcaption{25T}
    \end{subfigure}
    \begin{subfigure}{0.5\textwidth}
        \includegraphics[width = 4.5cm, height = 4cm]{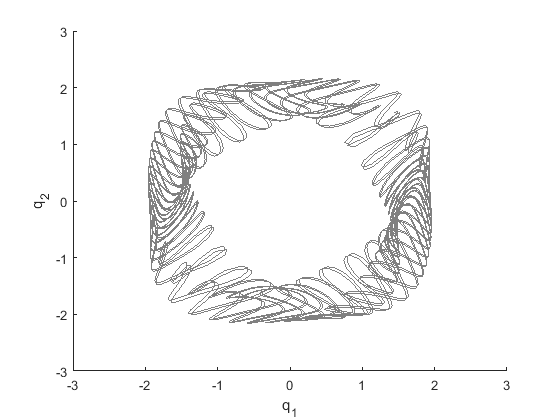}
        \subcaption{100T}
    \end{subfigure}
    \begin{subfigure}{0.5\textwidth}
        \includegraphics[width = 4.5cm, height = 4cm]{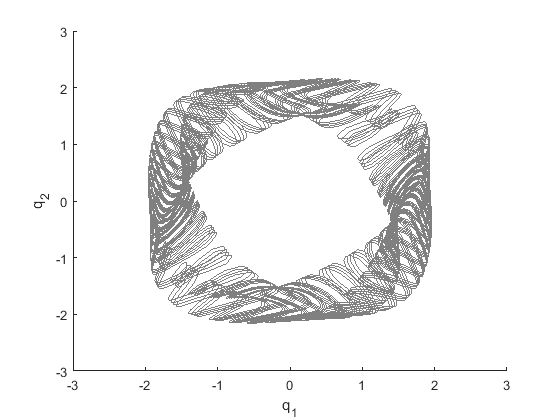}
        \subcaption{200T}
    \end{subfigure}
    \begin{subfigure}{0.5\textwidth}
        \includegraphics[width = 4.5cm, height = 4cm]{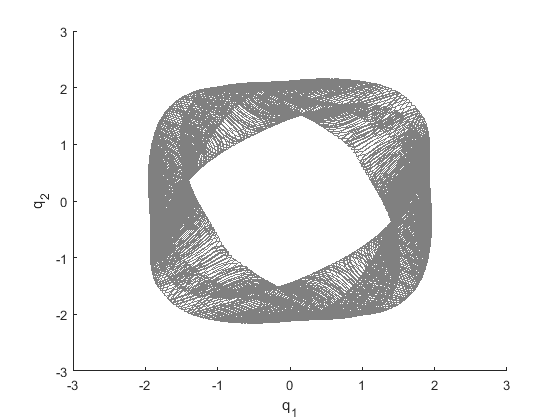}
        \subcaption{500T}
    \end{subfigure}
    \begin{subfigure}{0.5\textwidth}
        \includegraphics[width = 4.5cm, height = 4cm]{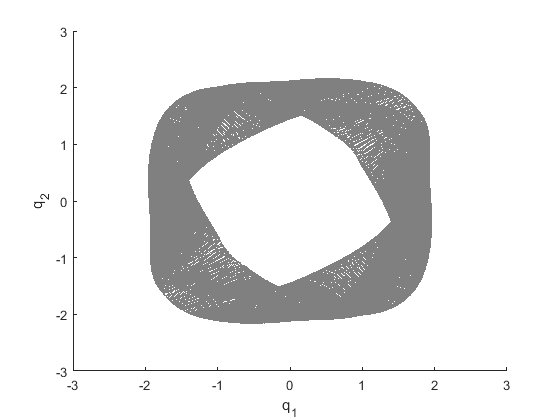}
        \subcaption{1000T}
    \end{subfigure}
    \begin{subfigure}{0.5\textwidth}
        \includegraphics[width = 4.5cm, height = 4cm]{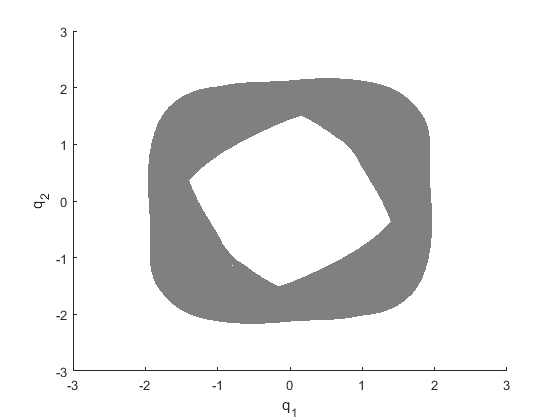}
        \subcaption{3000T}
    \end{subfigure}
     \caption{The trajectory in Fig. 7(a) plotted at times $25T$, $100T$, $200T$, $500T$, $1000T$ and $3000T$.}
\end{figure}
\pagebreak
When $\epsilon$ is decreased even further, to $0.1$ and then to $0.05$, the general pattern of behaviour found for $\epsilon = 0.25$ continues to hold.

For $\epsilon = 0.1$ the initial points 1-4 travel in ever narrower annular regions centered on the origin and with support at the edges of the main support of $|\psi|^2$, while points 5-10 are confined to very small regions (Fig. 9).

\begin{figure}[H]
    \begin{subfigure}{0.5\textwidth}
        \includegraphics[width = 6cm, height = 5cm]{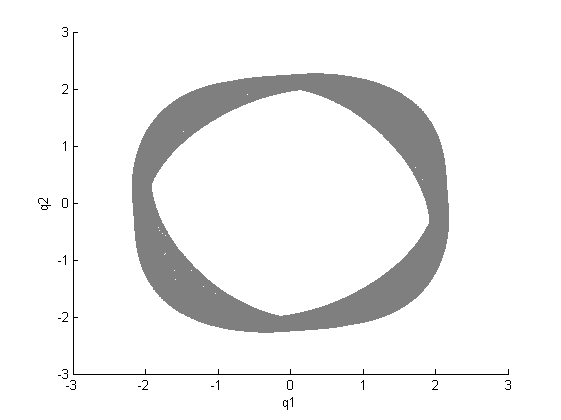}
        \subcaption{}
    \end{subfigure}
    \begin{subfigure}{0.5\textwidth}
        \includegraphics[width = 6cm, height = 5cm]{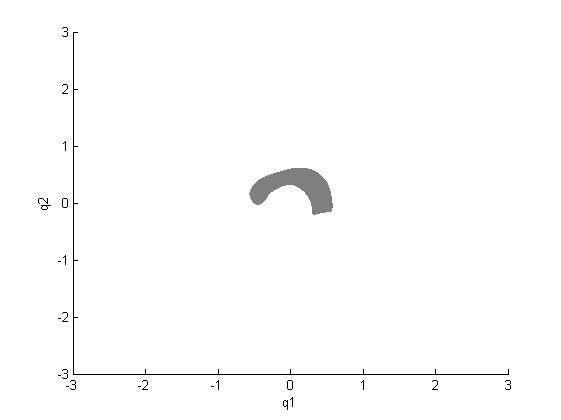}
        \subcaption{}
    \end{subfigure}
    \caption{Part (a) shows the trajectory starting from (-1.5, 1.5) plotted at the end of 3000 periods, for $\epsilon = 0.1$, with the set of initial phases listed in footnote 5. The trajectory travels over a narrow annular region centered on the origin. Part (b) shows the confined trajectory starting from (-0.5, 0) plotted at the end of 3000 periods, for the same wave function. We expect that this trajectory will also form a closed ring centered on the origin, after sufficiently long times.}
\end{figure}
\pagebreak
For $\epsilon = 0.05$ the inner and outer trajectories are confined to an even greater degree. Examples are shown in Fig. 10.
\begin{figure}[H]
    \begin{subfigure}{0.5\textwidth}
        \includegraphics[width = 6cm, height = 5cm]{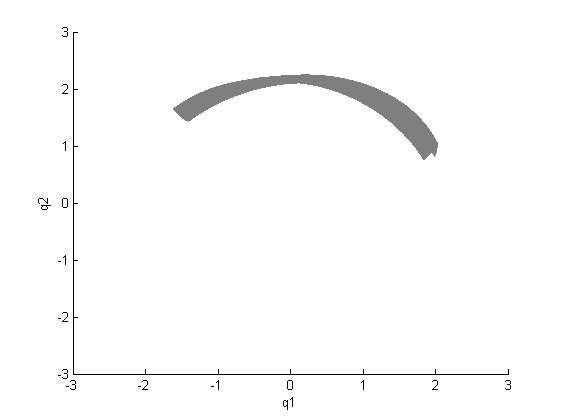}
        \subcaption{}
    \end{subfigure}
    \begin{subfigure}{0.5\textwidth}
        \includegraphics[width = 6cm, height = 5cm]{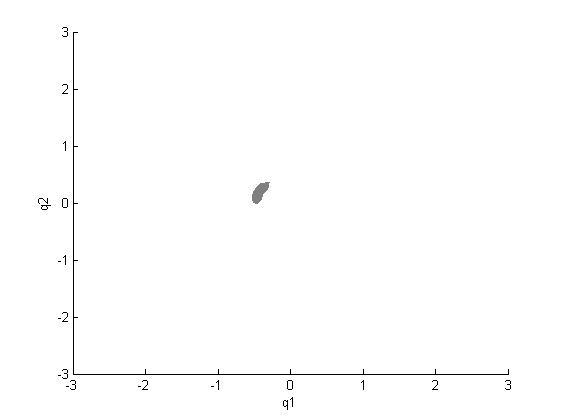}
        \subcaption{}
    \end{subfigure}
    \caption{Part (a) shows the trajectory starting from (-1.5, 1.5) plotted at the end of 3000 periods, for $\epsilon = 0.05$, with the set of initial phases listed in footnote 5. The trajectory is confined to a narrow region which will presumably form a ring over even longer timescales. Part (b) shows the highly confined trajectory starting from (-0.5, 0) plotted at the end of 3000 periods, for the same wave function.}
\end{figure}

For such small values of $\epsilon$, the outer trajectories are confined to annular regions centered on the origin - see for example Fig. 9(a). Close examination of the trajectories at intermediate times shows that they are confined to small regions with a superposed mean angular drift (as observed already in the example of Fig. 8, though with a somewhat narrower annular region). This is found to occur for all four outer trajectories for all sets of initial phases. The rate of angular drift is roughly the same for all the outer trajectories for a given set of initial phases -- but it varies depending on the choice of initial phases. While these trajectories travel over large distances, as far as relaxation is concerned they are highly confined.
\pagebreak

The above results are not specific to a homogeneous perturbation with $\epsilon_{mn} = \epsilon$. To confirm this, we may consider a specific inhomogeneous perturbation, for example with $\epsilon_{01} = 0.2$, $\epsilon_{10} = 0.15$ and $\epsilon_{11} = 0.1$. Illustrative examples of the trajectories for such a wave function are displayed in Fig. 11. The general behaviour is found to be comparable to that seen for homogeneous perturbations.

\begin{figure}[H]
    \begin{subfigure}{0.5\textwidth}
        \includegraphics[width = 6cm, height = 5cm]{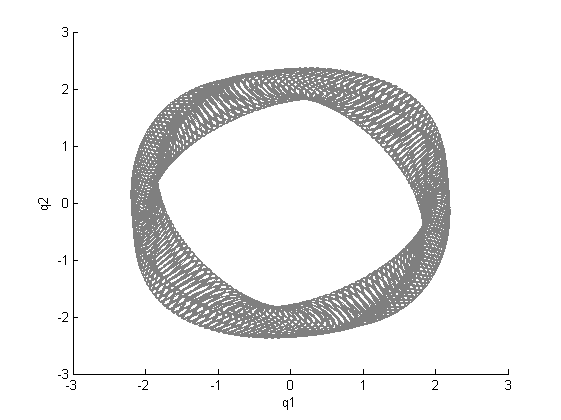}
        \subcaption{}
    \end{subfigure}
    \begin{subfigure}{0.5\textwidth}
        \includegraphics[width = 6cm, height = 5cm]{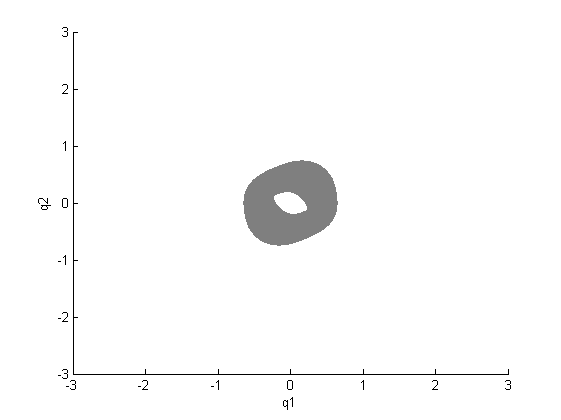}
        \subcaption{}
    \end{subfigure}
    \caption{Part (a) shows the trajectory starting from (-1.5, 1.5) plotted at the end of 3000 periods, for $\epsilon_{01} = 0.2$, $\epsilon_{10} = 0.15$ and $\epsilon_{11} = 0.1$, with the set of initial phases listed in footnote 5. The trajectory exhibits similar behaviour to that seen in Fig. 9(a). Part (b) shows the confined trajectory starting from (-0.5, 0) plotted at the end of 3000 periods, for the same wave function. This is comparable to the expected behaviour in Fig. 9(b) over sufficiently long timescales.}
\end{figure}

If we take the trajectories we have calculated (for many different sets of phases) as representative of the behaviour over very long timescales, we may draw a simple tentative conclusion. As the perturbation in the initial wave function is made smaller, the extent to which the trajectories explore the support of $|\psi|^2$ becomes smaller. We then expect that, for smaller perturbations, the extent of relaxation will be smaller (even in the long time limit). From our results, indeed, we may reasonably conclude that the extent of relaxation at large times will vanish for $\epsilon \rightarrow 0$.

\subsection{Small perturbations with two additional modes}

So far, we have considered wave functions with the ground state perturbed by a superposition of the $|1,0>$, $|0,1>$ and $|1,1>$ energy eigenstates of the two-dimensional harmonic oscillator. However, the $|2,0>$ and $|0,2>$ states have the same energy as the $|1,1>$ state and there is no particular reason to assume that they will be less likely to be present in a perturbation than the $|1,1>$ state. In this section we discuss the results of the simulations for perturbations with all five aforementioned excited states present, with the perturbation parameter $\epsilon = 0.1$.

We find that the trajectories are slightly less confined than they tend to be without the two newly added states (the annular regions are a bit thicker), but still do not come close to traveling over the bulk of the support of $|\psi|^2$. Thus we again find an absence of relaxation for small perturbations. Examples are shown in Fig. 12.\footnote{For the trajectories in Fig. 12, the initial phases were $\theta_{00} = 4.2065$, $\theta_{01} = 0.1803$, $\theta_{02} = 2.0226$, $\theta_{10} = 5.5521$, $\theta_{11} = 3.3361$, $\theta_{20} = 2.6561$.}

\begin{figure}[H]
    \begin{subfigure}{0.5\textwidth}
        \includegraphics[width = 6cm, height = 5cm]{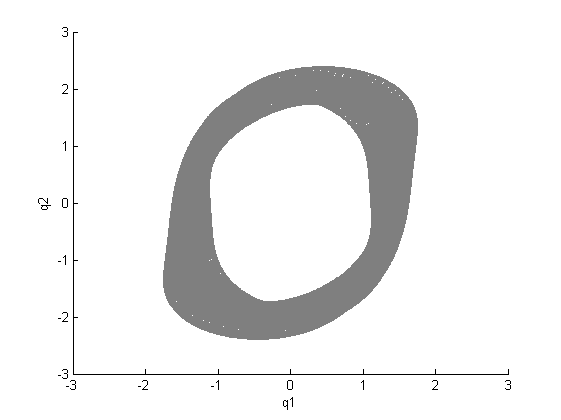}
        \subcaption{}
    \end{subfigure}
    \begin{subfigure}{0.5\textwidth}
        \includegraphics[width = 6cm, height = 5cm]{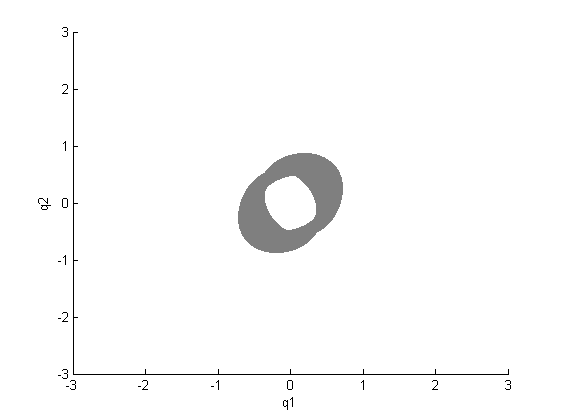}
        \subcaption{}
    \end{subfigure}
    \caption{Part (a) shows the trajectory starting from (1.5, 1.5) plotted at the end of 3000 periods, for $\epsilon=0.1$ and the initial phases specified in footnote 6, when the first five excited states are included in the perturbation. Part (b) shows the trajectory starting from (-0.5, 0) plotted at the end of 3000 periods, for the same wave function.}
\end{figure}

\subsection{Confinement of neighboring trajectories}

Despite the above results, it might be thought that relaxation could in principle still occur if neighboring initial points generated trajectories covering different sub-regions, in such a way that a small initial region explored the full support of $|\psi|^2$. As further evidence that relaxation does not occur, even over arbitrarily long timescales, we now show that in fact neighboring initial points generate trajectories that cover essentially the same sub-region of the support of $|\psi|^2$. As we shall see, neighboring initial points can generate trajectories that diverge widely, but even so the trajectories remain confined to the same sub-region. In such circumstances, relaxation cannot occur.

We consider 10 small squares of edge 0.04 centered at the points 1-10 listed above, with trajectories calculated for 13 initial points in each square.\footnote{For a square centered at the origin, the coordinates of the 13 points would be as follows: $(-0.02,0.02)$, $(-0.02,0.0)$, $(-0.02,-0.02)$, $(0.0,-0.02)$, $(0.02,-0.02)$, $(0.02,0.0)$, $(0.02,0.02)$, $(0.0,0.02)$, $(-0.01,0.01)$, $(-0.01,-0.01)$, $(0.01,-0.01)$, $(0.01,0.01)$, $(0.0,0.0)$.} This process was repeated for three different sets of phases. For these simulations we use the wave function with five homogeneous perturbative modes added to the ground state (as in section 4.3).

We find that the trajectories for all the points inside a given square explore almost exactly the same sub-region of the support of $|\psi|^2$. This is illustrated in Fig. 13.\footnote{For the trajectories displayed in Fig. 13, the initial phases used were $\theta_{00} = 1.2434$, $\theta_{01} = 4.411$, $\theta_{02} = 4.3749$, $\theta_{10} = 4.2427$, $\theta_{11} = 1.5574$, $\theta_{20} = 5.7796$.}

\begin{figure}[H]
    \begin{subfigure}{0.5\textwidth}
        \includegraphics[width = 6cm, height = 5cm]{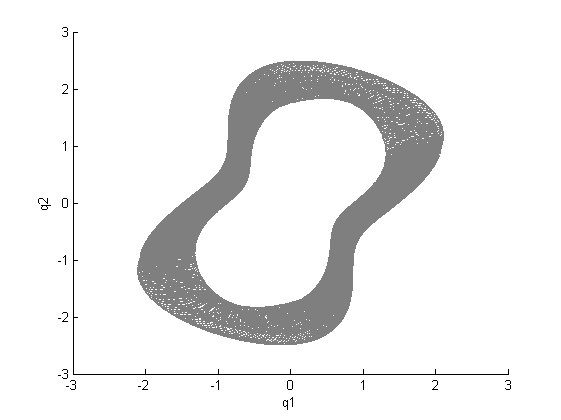}
        \subcaption{}
    \end{subfigure}
    \begin{subfigure}{0.5\textwidth}
        \includegraphics[width = 6cm, height = 5cm]{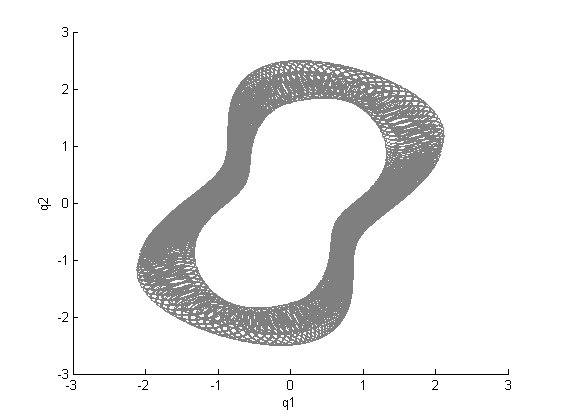}
        \subcaption{}
    \end{subfigure}
    \caption{Two trajectories starting from neighboring initial points in the small square centered at (1.5, 1.5), for the initial wave function with $\epsilon = 0.1$ and the phases specified in footnote 8. The trajectories cover almost exactly the same sub-region.}
\end{figure}
In some cases, however, a trajectory starting from one of the points in a square covered only a portion of the sub-region explored by its neighbors. An example is shown in Fig. 14.

\begin{figure}[H]
    \begin{subfigure}{0.5\textwidth}
        \includegraphics[width = 6cm, height = 5cm]{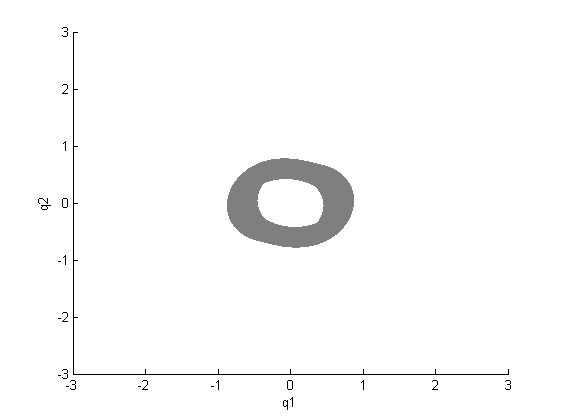}
        \subcaption{}
    \end{subfigure}
    \begin{subfigure}{0.5\textwidth}
        \includegraphics[width = 6cm, height = 5cm]{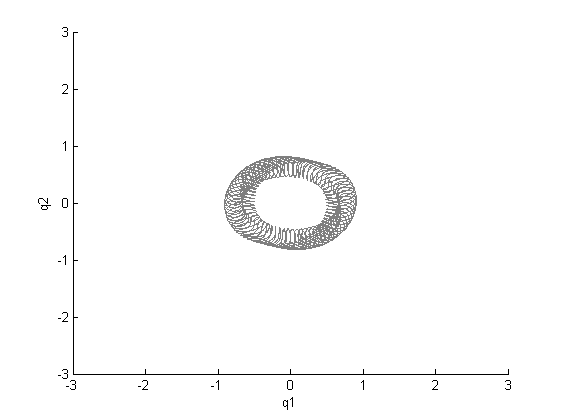}
        \subcaption{}
    \end{subfigure}
    \caption{Two trajectories starting from neighboring initial points in the small square centered at (0.5, 0.0), for the same wave function as in Fig. 14, at the end of 3000 periods. The trajectory in part (a) covers the sub-region more or less densely, while the trajectory in part (b) leaves significant empty spaces.}
\end{figure}

To ensure that our results are not specific to a homogeneous perturbation (with a common value of $\epsilon$), we have applied the same analysis for a wave function with the inhomogeneous perturbation parameters $\epsilon_{01} = 0.11$, $\epsilon_{02} = 0.12$, $\epsilon_{10} = 0.13$, $\epsilon_{11} = 0.14$ and $\epsilon_{20} = 0.15$. The results are similar. An example is displayed in Fig. 15.\footnote{For the trajectories displayed in Fig. 15, the initial phases used were $\theta_{00} = 4.0857$, $\theta_{01} = 0.2194$, $\theta_{02} = 4.6059$, $\theta_{10} = 1.2201$, $\theta_{11} = 0.439$, $\theta_{20} = 4.0563$.}

\begin{figure}[H]
    \begin{subfigure}{0.5\textwidth}
        \includegraphics[width = 6cm, height = 5cm]{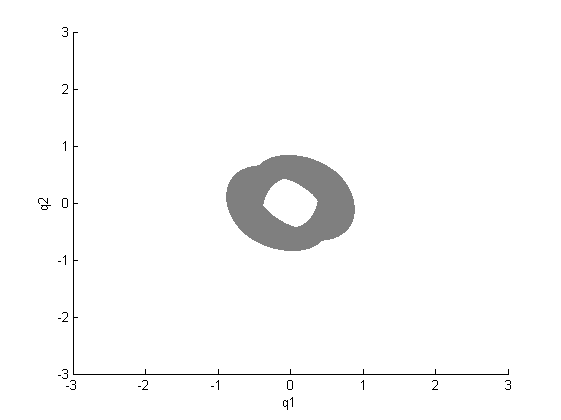}
        \subcaption{}
    \end{subfigure}
    \begin{subfigure}{0.5\textwidth}
        \includegraphics[width = 6cm, height = 5cm]{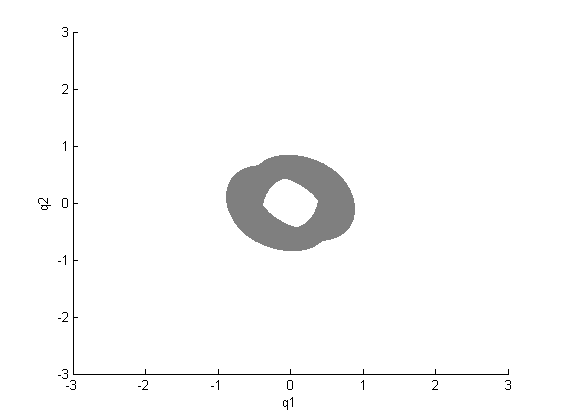}
        \subcaption{}
    \end{subfigure}
    \caption{Two trajectories starting from neighboring initial points in the small square centered at (0.25, 0.25), for the wave function with the perturbation amplitudes specified above and phases given in footnote 9, at the end of 3000 periods. The trajectories once again explore almost exactly the same sub-region.}
\end{figure}
It might be thought that the confinement of neighboring initial points to essentially the same evolved sub-region arises simply from the trajectories remaining close to each other. This is not always the case, however. Remarkably, some initial points within the same small square can become temporarily far apart (within the sub-region) and then very close again -- indicating that a confinement mechanism is at play. For example, when two of the trajectories in Fig. 13 were compared, the maximum distance between the positions over 3000 periods was 1.47 - which is roughly 15 percent of the diameter of the evolved sub-region. However, the distance between the final positions was only 0.08, which is of the same order of magnitude as the distance between the initial points.

\section{Conclusion}

Our numerical results provide evidence that small perturbations will not, in
fact, cause significant relaxation -- not even over arbitrarily long
timescales. In the examples we have studied, the system trajectories are confined to sub-regions of the support of $|\psi|^2$. Furthermore, neighboring initial points generate trajectories that are confined to essentially the same sub-regions. Such behavior precludes relaxation.

We have restricted ourselves to the harmonic oscillator, which as
we explained in Section 1 provides a testing ground applicable to high-energy
field theory in the early universe. In future work it would be of interest to
consider other systems, as well as to develop an analytical understanding of
the results (for which the methods of ref. \cite{CDE12} may prove useful). It has been suggested \cite{EC06, CDE12} that the extent and rate of relaxation might be related to the Lyapunov exponents of the trajectories. In the perturbative cases studied here the quantum relaxation is highly suppressed, while in the non-perturbative cases studied elsewhere the relaxation is usually rapid and very efficient \cite{VW05, EC06, TRV12, SC12, ACV14}. It would be interesting to evaluate Lyapunov exponents for these two extremes, as well as for intermediate cases, in order to explore the suggested relation. We leave this as a topic for future research.

One might question the extent to which our results for the two-dimensional oscillator can be relevant to cosmology. In fact, the formalism of cosmological perturbation theory shows that, in the relevant approximation, the effective Lagrangians of both scalar and tensor perturbations are equivalent to the Lagrangian of a free scalar field \cite{LL00, Muk05, PU09}. While one may question the motivation for choosing a particular initial quantum state, it is standard at least in inflationary cosmology to assume that the inflaton field is in the vacuum state. The Fourier modes are then unentangled and we have in effect a collection of independent harmonic oscillators. Non-vacuum states are not usually considered. An exception is ref. \cite{PLB99} which allows each mode to be in an arbitrary superposition of excited states, although with no entanglement between the modes. It might seem more logical to allow entangled states, but this will generally conflict with statistical homogeneity (which is well known to imply a diagonal two-point correlation function \cite{LL00, Muk05, PU09}). In applying our results to perturbations of the primordial vacuum, we are in effect considering small excitations of each mode but with no entanglement between modes. This is certainly simpler and perhaps even cosmologically preferred. In any case it makes our simulations tractable. In future work it would be interesting to consider the possible effect of entanglement between modes but this would require simulations in at least four dimensions.

From the point of view of a general understanding of relaxation, our results
suggest that in de Broglie-Bohm theory quantum equilibrium cannot be
understood as arising from the effects of small perturbations only, not even
in the long-time limit. Since all systems we know of have a long and violent
astrophysical history (ultimately stretching back to the big bang), their
current obedience to the Born rule may nevertheless be understood in terms of
the efficient relaxation found in previous simulations (at least at the
sub-Hubble wavelengths relevant to laboratory physics) for wave functions with
significant contributions from a range of energy states.

As regards cosmology, our results point to the following conclusions. Firstly,
the implicit assumption made in refs. \cite{AV07, AV10, CV13, CV15, AV15, CV16} is
justified: small corrections to the Bunch-Davies vacuum during inflation are
unlikely to cause significant relaxation, and so the derived predictions for
the CMB still stand (for the assumed scenario with a pre-inflationary period).
Secondly, relic nonequilibrium particles from the early universe surviving to
the present day remains a possibility at least in principle (albeit a rather
remote one for other reasons, as discussed in ref. \cite{UV15}).

\end{document}